\documentclass[a4paper,12pt]{article}  % other types include report, book, letter, slides

\usepackage{amsmath,amssymb}
\usepackage{setspace}
\usepackage[left=1.5cm,top=2cm,bottom=2.5cm,right=1.5cm,nohead]{geometry}
\usepackage{graphicx}
\usepackage{alltt}
\usepackage{listings}
\usepackage{dcolumn}
\usepackage{bm}
\usepackage{hyperref}
\usepackage{amsmath,amssymb}

\newcommand{\bt}[1]{{\mathbf #1}}

\renewcommand{\thefootnote}{\fnsymbol{footnote}}

\begin{document}
\title{\Large \textbf{Counting of discrete Rossby/drift wave resonant triads}}
\author{Miguel D. Bustamante${}^{1,}$\footnote{E-mail: \texttt{miguel.bustamante@ucd.ie}}
, Umar Hayat${}^2$, Peter Lynch${}^1$, Brenda Quinn${}^1$\\
{\small ${}^1$ School of Mathematical Sciences,
                University College Dublin,  Belfield, Dublin 4, Ireland}\\
{\small ${}^2$ Department of Mathematics, Quaid-i-Azam University,  Islamabad 45320, Pakistan}}

\maketitle

\bibliographystyle{plain}  % want neuron

\abstract{The purpose of this note is to remove the confusion about counting
of resonant wave triads for Rossby and drift waves in the context of the
Charney-Hasegawa-Mima equation.  In particular, we aim to point out a major error
of over-counting of triads
in the paper {\it Discrete exact and quasi-resonances of Rossby/drift waves
on $\beta$-plane with periodic boundary conditions}~\cite{[Kartashov and Kartashova 2013]}.

}

\section{Preliminaries}

\renewcommand{\thefootnote}{\arabic{footnote}}

The Charney-Hasegawa-Mima (CHM) equation on a bi-periodic domain $[0,2\pi)^2$ in physical
space for an infinite deformation radius is
 \begin{equation}
 \label{CHMx}
 \partial_t \Delta \psi + \beta \partial_x \psi +
 (\partial _x \psi) \partial_y \Delta \psi - (\partial _y \psi) \partial_x \Delta \psi = 0,
 \end{equation}
where $\psi=\psi(x,y,t)$ (a real-valued function) and $\beta$ is a constant. Let us introduce $\hat\psi_{\bt k},$ the Fourier transform of $\psi(x,y,t):$
\begin{equation}
\label{FourierTransform}
\psi(x,y,t) = \sum_{{\bt k} \in \mathbb{Z}^2} \hat\psi_{\bt k}(t) \exp(\mathrm{i} \,{\bt k}\cdot {\bt x}).
\end{equation}
The two-dimensional wavevectors are decomposed as ${\bt k} = (k_x, k_y).$ In the context of this discussion, we will restrict
the allowed interacting modes to those which are not zonal ($k_x \neq 0$) and, more importantly, we restrict the discussion to exactly resonant triads only: Fourier wavevectors ${\bt k}_1, {\bt k}_2, {\bt k}_3 \in \mathbb{Z}^2$ can interact if and only if
\begin{equation}
\label{reso}
{\bt k}_1 + {\bt k}_2 = {\bt k}_3 \quad \mathrm{and} \quad \omega({\bt k}_1)  + \omega({\bt k}_2) = \omega({\bt k}_3),
\end{equation}
where we introduced the linear dispersion $\omega({\bt k}) \equiv - \frac{\beta\,k_x}{|{\bt k}|^2}.$ The set of non-zonal wavevectors satisfying equations (\ref{reso}) is called ``resonant set'' and is denoted by $R (\subset \mathbb{Z}^2).$ In this case it is well established (see \cite{Piterbarg1988} and the book \cite{Nazarenko2011}, equations (7.8) and (6.11)) that the CHM equation can be cast in canonical form
 \begin{equation}
 \label{a_dot}
\mathrm{i} \,\dot{a}_{\bt k}= \omega_{\bt k} \,{a}_{\bt k}+ \mathrm{sign}(k_x) {\sum_{{\bt k}_1 + {\bt k}_2 = {\bt k}}} V_{1\, 2}^{{\bt k}} \,a_{{\bt k}_1} a_{{\bt k}_2} \,,
\end{equation}
with canonical variable\footnote{Equation (\ref{waveaction_a}) uses a slightly different notation as compared to equation (7.6) in \cite{Nazarenko2011}.}
\begin{equation}
\label{waveaction_a}
a_{\bt k} = -\frac{|{\bt k}|^2}{\sqrt{\beta |k_x|}}\hat{\psi}_{\bt k}\,.
\end{equation}
The nonlinear interaction coefficient is
\begin{equation}
\label{interactioncoeff}
V_{1\,2}^{{\bt k}} \equiv V_{{\bt k}_1, {\bt k}_2}^{{\bt k}} = \frac{\mathrm{i} \sqrt{\beta}}{2}\sqrt{|k_x k_{1x}k_{2x}|}\left(\frac{k_{1y}}{|{\bt k}_1|^2}+\frac{k_{2y}}{|{\bt k}_2|^2}- \frac{k_y}{|{\bt k}|^2}\right),
\end{equation}
and the sum in equation (\ref{a_dot}) is restricted to the resonant set $R.$ In particular, ${\bt k} \in R$  in this equation.

\section{Major error in \cite{[Kartashov and Kartashova 2013]}}

The over-counting problem in \cite{[Kartashov and Kartashova 2013]} is easily shown.
In equation (10) of that paper, six triads are listed. The authors state that
these are six different triads, but in fact all six are physically identical and mathematically equivalent.
The simplest way to see that there is redundancy is to compare the first
and fifth triad in equation (10): the only difference is that all wavenumbers
in the fifth triad are the negatives of the wavenumbers in the first.
But as is well known (see our equation (\ref{FourierTransform})), for a real field $\psi(x,y,t)$ the two modes $\hat{\psi}_{\bt k}$ and $\hat{\psi}_{-{\bt k}}$ must occur with amplitudes which are equal in absolute value. They are not independent.
 Thus, the claim of
\cite{[Kartashov and Kartashova 2013]} that they are listing six separate triads is wrong. %This six-fold degeneracy of triads is discussed explicitly in \cite{Bustamante2013}, equation (35)--(37).

In fact the problem is greater than this: all six triads are equivalent,
as we shall now show. \cite{[Kartashov and Kartashova 2013]} state that any exactly resonant triad ${{\bt k}_1 + {\bt k}_2={\bt k}_3}$ interacts with another five triads:
${\bt k}_1+(-{\bt k}_3)=(-{\bt k}_2),$ ${\bt k}_3+(-{\bt k}_2)={\bt k}_1,$
${\bt k}_3+(-{\bt k}_1)={\bt k}_2,$ $(-{\bt k}_2)+(-{\bt k}_1)=-{\bt k}_3$ and
${\bt k}_2+(-{\bt k}_3)=-{\bt k}_1.$
Crucially, in \cite{[Kartashov and Kartashova 2013]} the modes with wavenumbers ${\bt k}_1, {\bt k}_2, {\bt k}_3, (-{\bt k}_1), (-{\bt k}_2)$ and $(-{\bt k}_3)$ are considered
as \emph{six different modes}. In a general setting (complex-valued $\psi(x,y,t)$) this would be true because equation (\ref{a_dot}) alone gives separate evolution equations for
$a_{\bt k}$ and $a_{-\bt k}.$
However, due to the fact that
the underlying field $\psi(x,y,t)$ is \emph{real}, the extra condition
$a_{-\bt k}=\overline{a}_{\bt k}$ is imposed on the modes (overbars denote complex conjugation).
This extra condition is preserved in time
by the system of evolution equations (\ref{a_dot}),
allowing for a reduction of the original variables $\{a_{\bt k}\}_{{\bt k} \in R}$ to a smaller space
$\{a_{\bt k}\}_{{\bt k} \in \widetilde{R}}$ where $\widetilde{R}$ is the
restriction of the resonant set $R$ to the half plane $k_x > 0.$ Once this reduction is made, the
six triads stated in \cite{[Kartashov and Kartashova 2013]} as different reduce to a single
triad contributing with only one interaction term in the evolution equations. The six modes stated in \cite{[Kartashov and Kartashova 2013]} as different, reduce to three modes.\\

\noindent \textbf{Detailed reduction.} Let us look at the contributions from equations (\ref{a_dot}) to the evolution of the six modes' amplitudes ${a}_{{\bt k}_1}, {a}_{{\bt k}_2}, {a}_{{\bt k}_3}, {a}_{-{\bt k}_1}, {a}_{-{\bt k}_2}, {a}_{-{\bt k}_3},$ \emph{without using the extra condition}
$a_{-\bt k}=\overline{a}_{\bt k}.$ We obtain
\begin{eqnarray}
\nonumber
\mathrm{i} \,\dot{a}_{{\bt k}_1} &=& \omega_{{\bt k}_1} \,{a}_{{\bt k}_1} + 2\, \mathrm{sign}(k_{1x}) \,  V_{-{\bt k}_2, {\bt k}_3}^{{\bt k}_1} \, {a}_{-{\bt k}_2} a_{{\bt k}_3}  + \ldots\,,\\
\nonumber
\mathrm{i} \,\dot{a}_{-{\bt k}_2} &=& \omega_{-{\bt k}_2} \,{a}_{-{\bt k}_2} + 2\, \mathrm{sign}(-k_{2x}) \,V_{-{\bt k}_3, {\bt k}_1}^{-{\bt k}_2} \,a_{-{\bt k}_3} {a}_{{\bt k}_1} + \ldots\,,\\
\nonumber
\mathrm{i} \,\dot{a}_{{\bt k}_3} &=& \omega_{{\bt k}_3} \,{a}_{{\bt k}_3} + 2\, \mathrm{sign}(k_{3x}) \, V_{{\bt k}_1, {\bt k}_2}^{{\bt k}_3} \,a_{{\bt k}_1} {a}_{{\bt k}_2} + \ldots\,,\\
\nonumber
\mathrm{i} \,\dot{a}_{-{\bt k}_3} &=& \omega_{-{\bt k}_3} \,{a}_{-{\bt k}_3} + 2\, \mathrm{sign}(-k_{3x}) \,V_{-{\bt k}_1, -{\bt k}_2}^{-{\bt k}_3} \,a_{-{\bt k}_1} {a}_{-{\bt k}_2} + \ldots\,,\\
\nonumber
\mathrm{i} \,\dot{a}_{{\bt k}_2} &=& \omega_{{\bt k}_2} \,{a}_{{\bt k}_2} + 2\, \mathrm{sign}(k_{2x}) \,V_{{\bt k}_3, -{\bt k}_1}^{{\bt k}_2} \,a_{{\bt k}_3} {a}_{-{\bt k}_1} + \ldots\,,\\
\label{triadeqns}
\mathrm{i} \,\dot{a}_{-{\bt k}_1} &=& \omega_{-{\bt k}_1} \,{a}_{-{\bt k}_1} + 2\, \mathrm{sign}(-k_{1x}) \,V_{{\bt k}_2, -{\bt k}_3}^{-{\bt k}_1} \,{a}_{{\bt k}_2}  a_{-{\bt k}_3} + \ldots\,,
\end{eqnarray}
where ``$\ldots$'' denote terms involving modes beyond ${\bt k}_1, {\bt k}_2, {\bt k}_3, (-{\bt k}_1), (-{\bt k}_2)$ and $(-{\bt k}_3).$ To derive these equations we used the symmetry of the interaction coefficient $V_{{\bt k}, {\bt k}'}^{{\bt k}''} = V_{{\bt k}', {\bt k}}^{{\bt k}''}$ stemming from equation (\ref{interactioncoeff}).

Equations (\ref{triadeqns}) appear like a system of six coupled equations for six complex variables. However, once the extra condition $a_{-{\bt k}} = \overline{a}_{\bt k}$ (stemming from the reality of the underlying field $\psi(x,y,t)$) is used, these six equations reduce to only three independent equations, for three independent variables $a_{{\bt k}_1}, a_{{\bt k}_2}, a_{{\bt k}_3}.$ We show explicitly that the first and the last equations in (\ref{triadeqns}) are equivalent. The remaining pairings can be shown in a similar way. The complex conjugate of the first equation gives
$$- \mathrm{i} \,\dot{\overline{a}}_{{\bt k}_1} = \omega_{{\bt k}_1} \,\overline{a}_{{\bt k}_1} + 2\, \mathrm{sign}(k_{1x}) \,\overline{V}_{-{\bt k}_2, {\bt k}_3}^{{\bt k}_1} \, \overline{a}_{-{\bt k}_2} \overline{a}_{{\bt k}_3}  + \ldots \, .
$$
Using $\overline{a}_{\bt k} = a_{-{\bt k}}$ and $\overline{V}_{-{\bt k}_2, {\bt k}_3}^{{\bt k}_1} = -{V}_{-{\bt k}_2, {\bt k}_3}^{{\bt k}_1} = {V}_{{\bt k}_2, -{\bt k}_3}^{-{\bt k}_1}$ we obtain
$$- \mathrm{i} \,\dot{{a}}_{-{\bt k}_1} = \omega_{{\bt k}_1} \,{a}_{-{\bt k}_1} + 2\, \mathrm{sign}(k_{1x}) \,{V}_{{\bt k}_2, -{\bt k}_3}^{-{\bt k}_1} \, {a}_{{\bt k}_2} {a}_{-{\bt k}_3}  + \ldots\,,
$$
but this is equivalent to the last equation in (\ref{triadeqns}) after using the identity $\omega_{{\bt k}} = - \omega_{-{\bt k}}.$

In summary, equations (\ref{triadeqns}) reduce to the well-known triad system
\begin{eqnarray}
\nonumber
\mathrm{i} \,\dot{a}_{{\bt k}_1} &=& \omega_{{\bt k}_1} \,{a}_{{\bt k}_1} + 2\, \mathrm{sign}(k_{1x}) \,V_{{\bt k}_3, -{\bt k}_2}^{{\bt k}_1} \,a_{{\bt k}_3} \overline{a}_{{\bt k}_2} + \ldots \,,\\
\nonumber
\mathrm{i} \,\dot{a}_{{\bt k}_2} &=& \omega_{{\bt k}_2} \,{a}_{{\bt k}_2} + 2\, \mathrm{sign}(k_{2x}) \, V_{{\bt k}_3, -{\bt k}_1}^{{\bt k}_2} \,a_{{\bt k}_3} \overline{a}_{{\bt k}_1} + \ldots \,,\\
\label{triadeqnsReduced}
\mathrm{i} \,\dot{a}_{{\bt k}_3} &=& \omega_{{\bt k}_3} \,{a}_{{\bt k}_3} + 2\, \mathrm{sign}(k_{3x}) \, V_{{\bt k}_1, {\bt k}_2}^{{\bt k}_3} \,a_{{\bt k}_1} {a}_{{\bt k}_2} + \ldots  \,.
\end{eqnarray}

Therefore all six triads stated in \cite{[Kartashov and Kartashova 2013]} as different are in fact only one physical triad.

For the domain defined by $0 \leq |k_x|, |k_y| \leq 200,$
\cite{[Kartashov and Kartashova 2013]}
report finding $828$ triads while
\cite{Bustamante2013}
find a significantly lower number. The discrepancy is easily explained:
since \cite{[Kartashov and Kartashova 2013]} over-count by a factor of 6, they have in reality
identified only $(828/6) = 138$ valid triads.
\cite{Bustamante2013} confined their search to the \emph{interior} of the domain $0 \leq |k_x|, |k_y| \leq 200$ and found $136$ triads.
Two additional triads have a vertex sitting on the boundary of the domain: $(24,88), (40,-200),(64,-112)$ and $(24,-88), (40,200),(64,112).$
This gives a total of $138$ triads, precisely one sixth of the number reported by
\cite{[Kartashov and Kartashova 2013]}.

\section{False deduction by \cite{[Kartashov and Kartashova 2013]} from a Theorem in \cite{[Yamada and Yoneda 2013]}}

In \cite{[Kartashov and Kartashova 2013]} it is stated that there is a contradiction between the work of \cite{Bustamante2013} and a mathematical theorem
in \cite{[Yamada and Yoneda 2013]}. However, \cite{[Yamada and Yoneda 2013]} considered the limit of large $\beta,$ whereas the results of \cite{Bustamante2013} are for finite $\beta.$
Moreover, the Theorem of \cite{[Yamada and Yoneda 2013]} relies on the presence of viscosity. \cite{[Kartashov and Kartashova 2013]} have drawn an
unjustified conclusion from the theorem in \cite{[Yamada and Yoneda 2013]}.

\section{Quasi-resonances}

\cite{[Kartashov and Kartashova 2013]} state that all quasi-resonances found in \cite{Bustamante2013} in a given domain \mbox{$0 \leq |k_x|, |k_y| \leq L$}
are formed in the neighborhood of exact resonant triads.
This is false. In fact, the majority of quasi-resonant triads found by the method of \cite{Bustamante2013} are arbitrarily far from the exact resonant triads
that are used to generate them, because these exact resonant triads are typically of wavenumbers much greater than $L$ (by orders of magnitude).
 More importantly, \cite{Bustamante2013} produce triads close to the so-called resonant manifold,
which is actually a set of curves on $\mathbb{R}^2,$ typically with only a few integer points, if any. These integer points correspond to exact resonances.
Incidentally, \cite{[Kartashov and Kartashova 2013]} attempt a similar method in their Section IV.

\section{Conclusions}

There are a number of other errors in \cite{[Kartashov and Kartashova 2013]}. However, the most egregious error is the $6$-fold over-counting of triads.

\section{Acknowledgments}

We thank G.~McGuire and S.~V.~Nazarenko for useful comments and suggestions.

%\bibliography{bibliography}

\end{document}